\documentclass[twocolumn,10pt]{article}
\usepackage[top=1in, bottom=1in, left=0.75in, right=0.75in]{geometry}
\usepackage{amsmath}
\usepackage{graphicx}
\usepackage{titlesec}
\usepackage{times}
\usepackage{authblk}
\usepackage{subcaption}
\usepackage{booktabs}
\usepackage[numbers, sort&compress]{natbib}

\usepackage[colorlinks=true, linkcolor=blue, citecolor=blue, urlcolor=blue]{hyperref}
\usepackage{subcaption} 
\titleformat{\section}{\large\bfseries}{\thesection}{1em}{}
\titleformat{\subsection}{\normalsize\bfseries}{\thesubsection}{1em}{}

\title{Explainable AI for Curie Temperature Prediction in Magnetic Materials}
\author[1]{M. Adeel Ajaib}
\affil[1]{\textit{Department of Data Sciences, Penn State Abington, Abington, PA 19001, USA} }
\author[2]{Fariha Nasir}
\affil[2]{\textit{Department of Physics, Penn State Abington, Abington, PA 19001, USA} }
\author[3]{ Abdul Rehman}
\affil[3]{\textit{Department of Physics,University of Delaware, Newark, DE 19716, USA}}
\date{}

\begin{document}

\twocolumn[
\maketitle
\begin{@twocolumnfalse}
\begin{abstract}
We explore machine learning techniques for predicting Curie temperatures of magnetic materials using the NEMAD database. 
By augmenting the dataset with composition-based and domain-aware descriptors, we evaluate the performance of several machine learning models. We find that the Extra Trees Regressor delivers the best performance, reaching an R\textsuperscript{2} score of up to 0.85 $\pm$ 0.01 (cross-validated) for a balanced dataset. We employ the k-means clustering algorithm to gain insights into the performance of chemically distinct material groups. Furthermore, we perform SHAP analysis to identify key physicochemical drivers of Curie behavior, such as average atomic number and magnetic moment. 
By employing explainable AI techniques, this analysis offers insights into the model’s predictive behavior, thereby advancing scientific interpretability.

\vspace{5mm}

\end{abstract}
\end{@twocolumnfalse}
]

\begin{figure*}[t]
    \centering
    \begin{minipage}[t]{0.48\textwidth}
        \centering
        \includegraphics[width=\linewidth]{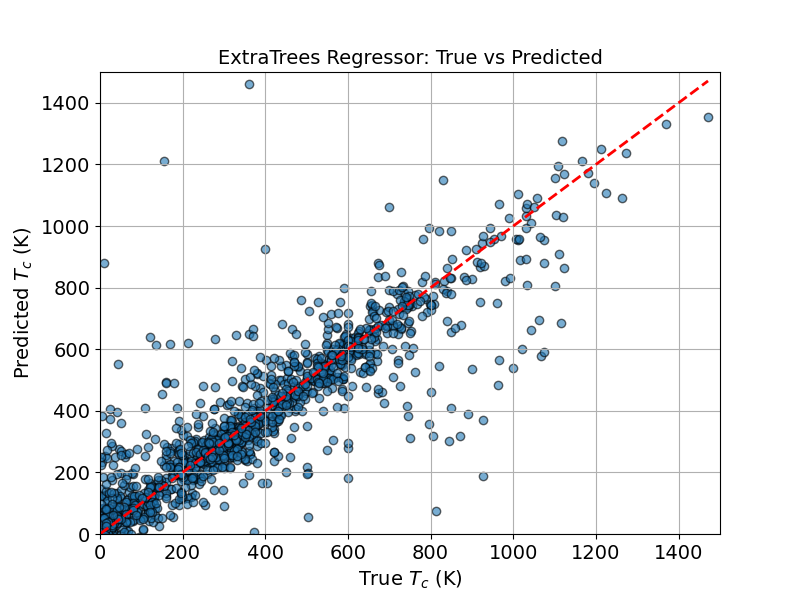}
        \caption*{(a) Main dataset}
    \end{minipage}%
    \hfill
    \begin{minipage}[t]{0.48\textwidth}
        \centering
        \includegraphics[width=\linewidth]{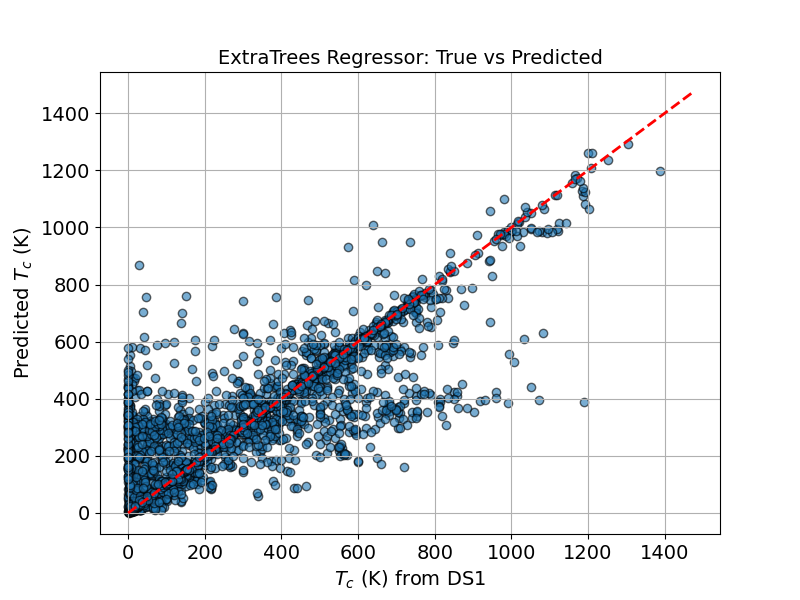}
        \caption*{(b) DS1 validation dataset}
    \end{minipage}
    \caption{Predicted vs. true Curie temperatures using the Extra Trees model for (a) the training/test dataset and (b) the DS1 validation dataset. A fairly strong alignment along the diagonal indicates the predictive performance of the model.}
    \label{fig:combined_true_vs_pred}
\end{figure*}

\begin{figure}[ht]
    \centering
    \includegraphics[width=0.45\textwidth]{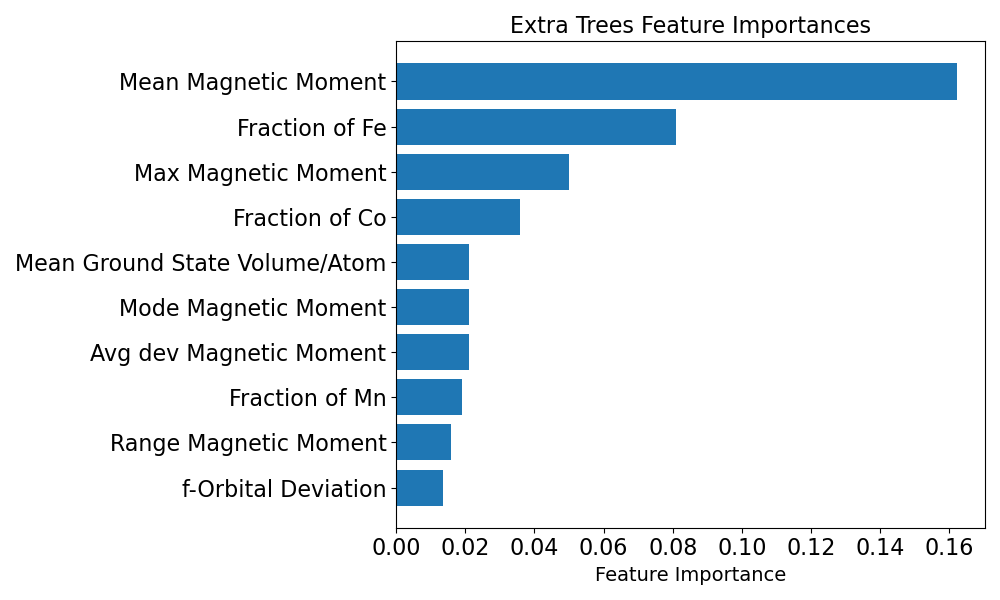}
    \caption{Feature importance graph for the Extra Trees Algorithm.}
    \label{fig:elbow_plot}
\end{figure}

\section{Introduction}
Curie temperature ($T_c$) is a critical property that defines the transition of a material from a ferromagnetic to a paramagnetic state. Predicting $T_c$ with high accuracy is essential for the discovery of magnetic materials for data storage, spintronics, and energy applications. Traditional approaches based on quantum mechanical computations or empirical models are often limited in scalability and accuracy. In recent years, machine learning (ML) has emerged as a promising alternative for property prediction across materials science domains \cite{rajan2015materials, jung2024, lu2022, brannvall2024, ward2016general, merker2022mlmag, schmidt2019recent, orang2025predicting, Ponomarova2025MLCurie}.

Building on this momentum, several recent studies have proposed the use of ML models trained on curated magnetic datasets. In particular, the recent study \cite{itani2024northeast} introduced the NEMAD database, which aggregates experimentally measured magnetic transition temperatures and compositions. 
Similarly, the study by ~\cite{belot2023systematic} utilized two of the largest available datasets of experimental Curie temperatures—comprising over 2,500 materials for training and more than 3,000 entries for validation—to compare machine learning strategies for predicting Curie temperature solely from chemical composition. 

Our work is inspired by these prior efforts and aims to improve the predictive accuracy and gain insights into model interpretability. We develop a pipeline that starts from the NEMAD dataset, augments it with compositional and elemental features, and evaluates several ML models. A key contribution of our work is the integration of explainable AI (XAI) through SHAP  (SHapley Additive exPlanations) analysis, which allows us to quantify how each input feature contributes to the model’s prediction. Moreover, we benchmark our models on external datasets   from literature to demonstrate generalization.

\section{Dataset Preparation and Model Evaluation}
We begin our study with the NEMAD database \cite{itani2024northeast}, a publicly available dataset compiled using large language models, comprising 26,706 magnetic materials with records for chemical composition, phase transition temperatures (e.g., Curie or Néel temperatures) and related magnetic properties. We preprocessed the database to remove ambiguous entries, duplicates, and entries with missing or non-numeric temperatures. We also standardized chemical formulas using the {pymatgen} library.

{From the original 33,668 entries in the NEMAD dataset, we retained 21,024 samples (62.4\%). Rows with missing material names or Curie temperatures (36.6\%) were removed. Curie temperatures were parsed to extract numerical Kelvin values, and entries with non-parsable or non-physical temperatures (approximately 1.5\% of the data) were discarded. This generated a clean dataset for the machine learning analysis.}

To generate descriptors, we first computed composition-based features such as average atomic number, atomic mass, group number, period, and electronegativity using {pymatgen}. We then added elemental property descriptors using the Magpie feature set via {matminer}, which includes statistics such as mean, range, and standard deviation of valence electron counts, ionic radii, and oxidation states. These features capture the diversity and bonding characteristics of constituent elements and have been shown to correlate with magnetic properties.

We evaluated several machine learning models for predicting Curie temperatures using the enriched dataset. Table~\ref{tab:model_performance} presents the comparative performance metrics in terms of Mean Absolute Error (MAE), Root Mean Squared Error (RMSE), and the coefficient of determination ($R^2$).

\begin{table}[h]
\centering
\caption{Performance comparison of machine learning models for Curie temperature prediction after feature normalization where necessary.}
\begin{tabular}{|l|c|c|c|}
\hline
\textbf{Model} & \textbf{MAE (K)} & \textbf{RMSE (K)} & \textbf{R\textsuperscript{2}} \\
\hline
Extra Trees       & 51 & 98  & 0.87 \\
Random Forest     & 57 & 103 & 0.85 \\
XGBoost           & 63 & 106 & 0.85 \\
Neural Network    & 69 & 123 & 0.79 \\
KNN               & 74 & 129 & 0.77 \\
\hline
\end{tabular}
\label{tab:model_performance}
\end{table}

The Extra Trees algorithm achieved the best overall performance, with an $R^2$ value of 0.87, MAE of 54 K, and RMSE of 106 K. This performance surpasses that of the Random Forest algorithm and other models considered. 
With a balanced dataset, we were able to achieve $R^2 \sim 0.85$ with a standard deviation of $0.01$. The balanced dataset was obtained by randomly removing around 4\% of samples with Curie temperatures less than 300 K  \cite{itani2024northeast}.

The Extra Trees (Extremely Randomized Trees) algorithm employs an ensemble learning method similar to the Random Forest model, but with two notable  distinctions:
\begin{itemize}
    \item It selects cut-points at random rather than optimizing splits.
    \item It uses the entire training dataset without bootstrapping.
\end{itemize}
These differences introduce greater variance among the individual trees in the ensemble, which can improve generalization and reduce overfitting on noisy datasets. This may explain its superior performance in our analysis, where the feature space is high-dimensional and the temperature distribution is non-uniform.

The Random Forest model also performs well but slightly underperforms Extra Trees, likely due to its more deterministic node-splitting strategy, which may be less adaptive to irregularities in the dataset.

XGBoost, KNN, and neural networks were also tested. XGBoost shows competitive results but did not outperform tree-based ensembles. KNN and neural networks showed relatively weaker performance, potentially due to the complexity of the feature relationships and sensitivity to scaling.

We validate the performance of the Extra Trees model on the DS1 dataset \cite{nelson2019predicting}. Figure~\ref{fig:combined_true_vs_pred} presents the predicted versus actual Curie temperatures using the Extra Trees model, both for the main dataset and the DS1 validation test dataset. In both subfigures, the data points cluster along the $y=x$ diagonal, indicating that the predicted and actual values agree fairly well ($R^2 \sim 0.71$, MAE $\sim$ 91 K) which demonstrates the model's generalizability.

\begin{figure*}[t]
    \centering
    \begin{minipage}[t]{0.48\textwidth}
        \centering
        \includegraphics[width=\linewidth]{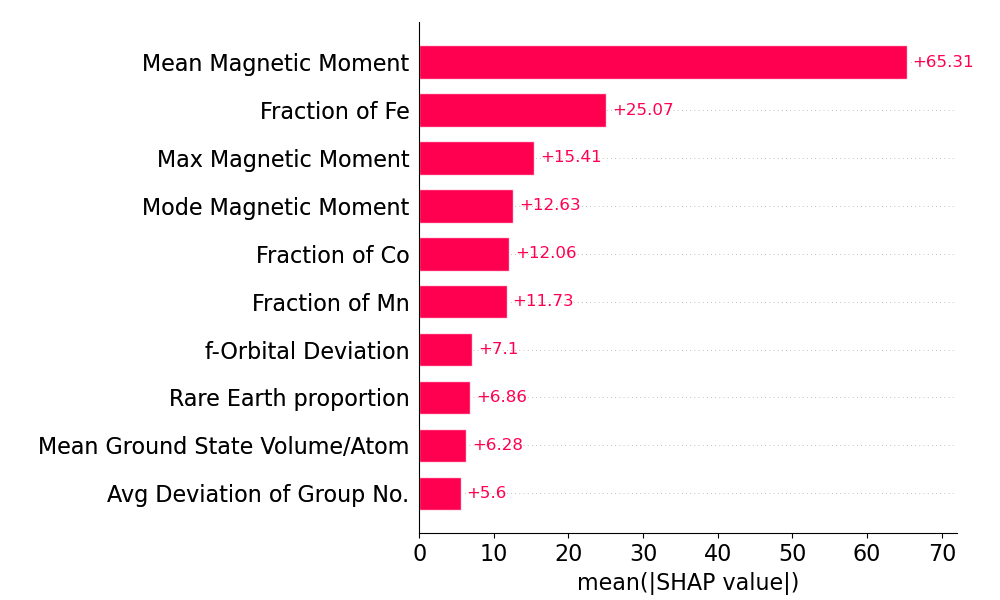}
        \caption*{(a) Cluster-wise elemental composition percentages.}
        \label{fig:shapbar}
    \end{minipage}%
    \hfill
    \begin{minipage}[t]{0.48\textwidth}
        \centering
        \includegraphics[width=\linewidth]{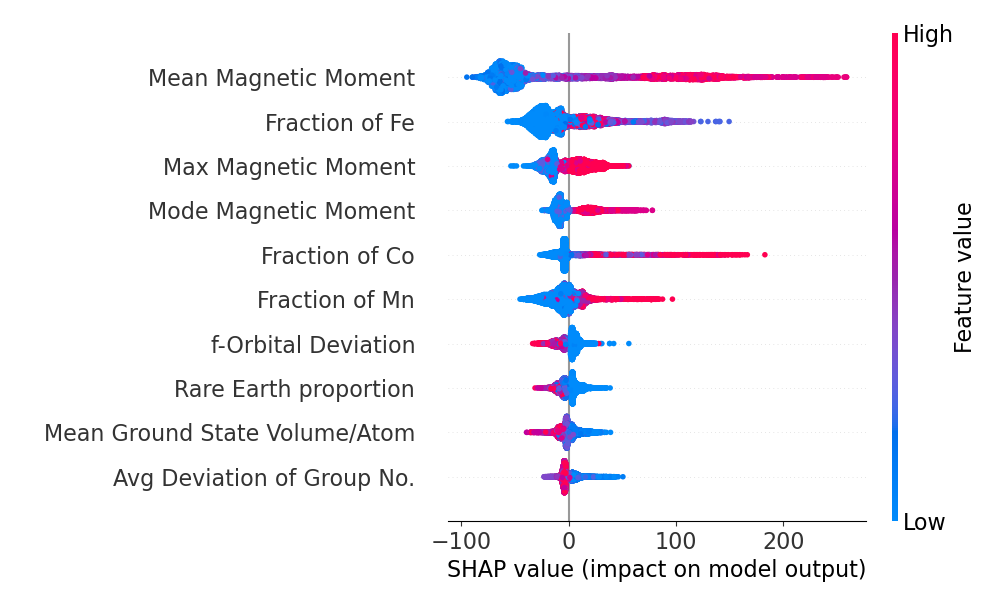}
        \caption*{(b) t-SNE projection with cluster labels.}
        \label{fig:shapbeeswarm}
    \end{minipage}
    \caption{SHAP analysis of the Extra Trees model. {Panel (a) shows SHAP bar plot showing the mean absolute SHAP value of the top features contributing to Curie temperature predictions across the entire dataset. Higher bars indicate stronger global influence on model output.} {Panel (b) displays the SHAP beeswarm plot illustrating the distribution of feature impacts for individual samples. Each point represents a material. Feature strength is encoded in color (red = high, blue = low), showing whether large or small values increase or decrease the predicted Curie temperature.}}
    \label{fig:shap_combined}
\end{figure*}

\subsection{SHAP Analysis}

To better understand the model’s behavior and identify the most influential features contributing to the prediction of Curie temperature, we performed SHAP (SHapley Additive exPlanations) analysis. SHAP values offer a model-agnostic, consistent way to explain predictions by attributing importance scores to individual features.

\textbf{SHAP Bar Plot:} Figure~\ref{fig:shap_combined} shows the SHAP summary bar plot of the top 10 features based on their mean absolute SHAP values. Among the dominant predictors, the {mean magnetic moment} stands out, reinforcing the well-established role of atomic-scale magnetism in determining $T_c$. The {fraction of Fe} also shows high importance, which aligns with its prevalence in high-$T_c$ compounds and well-known ferromagnets. Features such as the {mode magnetic moment}, {max magnetic moment}, and {fraction of Co} all rank highly, further confirming that the magnetic nature of the constituent atoms is a primary driver of the model’s predictions.

Interestingly, features such as {rare earth proportion}, {mean ground state volume per atom}, and {group number deviation} appear in the top contributors. These descriptors capture the complexity of atomic arrangements and compositional heterogeneity, both of which are known to modulate magnetic interactions in complex systems.

\textbf{SHAP Beeswarm Plot:} The SHAP beeswarm plot in  Figure~\ref{fig:shap_combined}, right panel, illustrates the distribution of SHAP values for each feature across all samples. Each point represents a single material, with color indicating the corresponding feature value. This visualization reveals not only which features are most influential, but also how they affect the model’s predictions. For example, we can see from the plot that the Mean Magnetic Moment, which represents the average magnetic moment of the ground-state structure, exhibits a strong positive influence when its value is high (shown in red), shifting SHAP values to the right. This suggests that materials with higher magnetic moments are strongly associated with higher Curie temperatures, which is physically intuitive.

In contrast, the {f-Orbital Deviation} shows a somewhat different pattern. High values (in red) are generally associated with negative SHAP values, indicating a suppressive effect on Curie temperature. This deviation metric captures the compositional spread in elements possessing f-electrons, such as rare-earth or actinide elements.

Taken together, these plots indicate that the model captures meaningful physical trends: magnetic moment-related features dominate, while elemental diversity exhibits subtle patterns. This supports the model’s reliability and its alignment with known physics.

\begin{figure}[ht]
    \centering
    \includegraphics[scale=0.33]{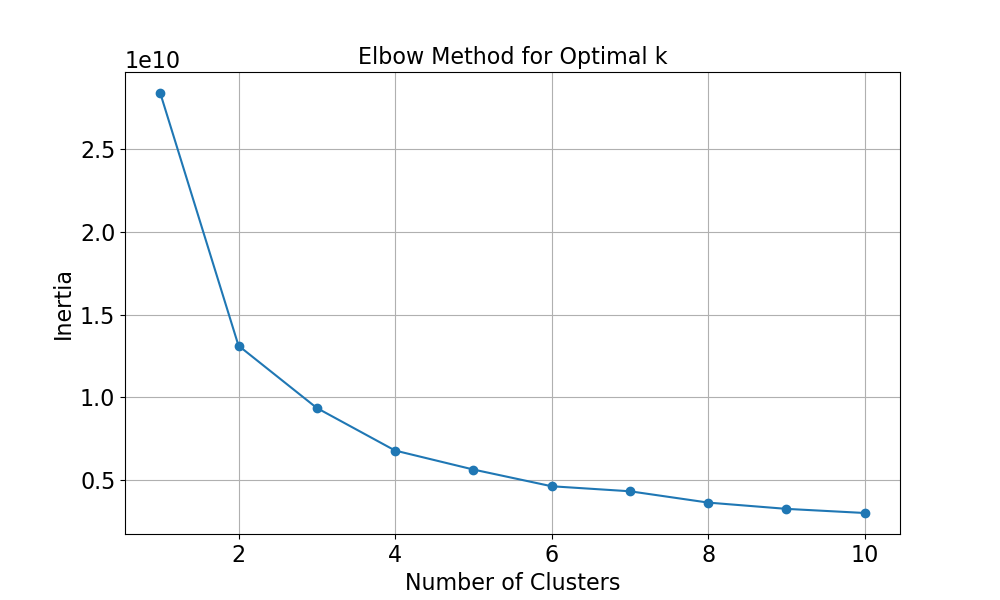}
    \caption{{Elbow plot showing the inertia values for K-means clustering as a function of the number of clusters $k$. 
A rapid decrease in inertia is observed from $k = 1$ to $k = 3$, followed by a clear flattening beyond $k = 4$, 
indicating that adding more clusters yields only marginal improvement in explaining the data structure. This behavior supports the choice of 
$k = 4$ as a balanced and physically meaningful clustering configuration for the Curie temperature dataset.}}
    \label{fig:elbow_plot}
\end{figure}

\section{K-means Clustering Analysis}
In this section, we present our implementation of the k-means clustering algorithm to identify subgroups within the dataset that may exhibit distinct magnetic behavior. To determine the optimal number of clusters, we employed the elbow method, as shown in Figure~\ref{fig:elbow_plot}. The curve shows a sharp drop in inertia from $k=1$ to $k=3$, with diminishing returns beyond $k=4$. This led us to select $k=4$ for clustering.

\begin{figure*}[t]
    \centering
    \begin{minipage}[t]{0.5\textwidth}
        \centering
        \includegraphics[width=\linewidth]{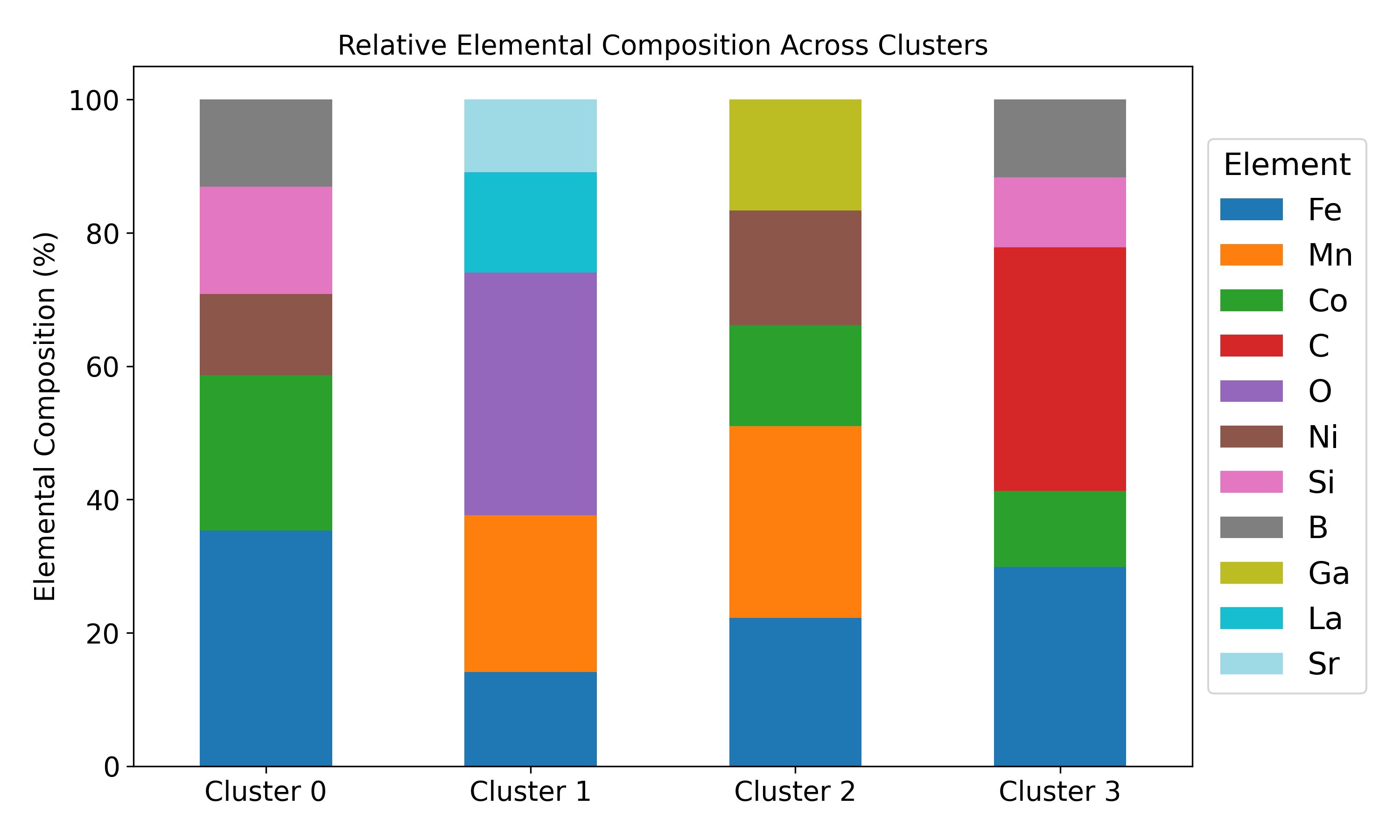}
        \caption*{(a) Cluster-wise elemental composition percentages}
    \end{minipage}
    \hfill
    \begin{minipage}[t]{0.48\textwidth}
        \centering
        \includegraphics[width=\linewidth]{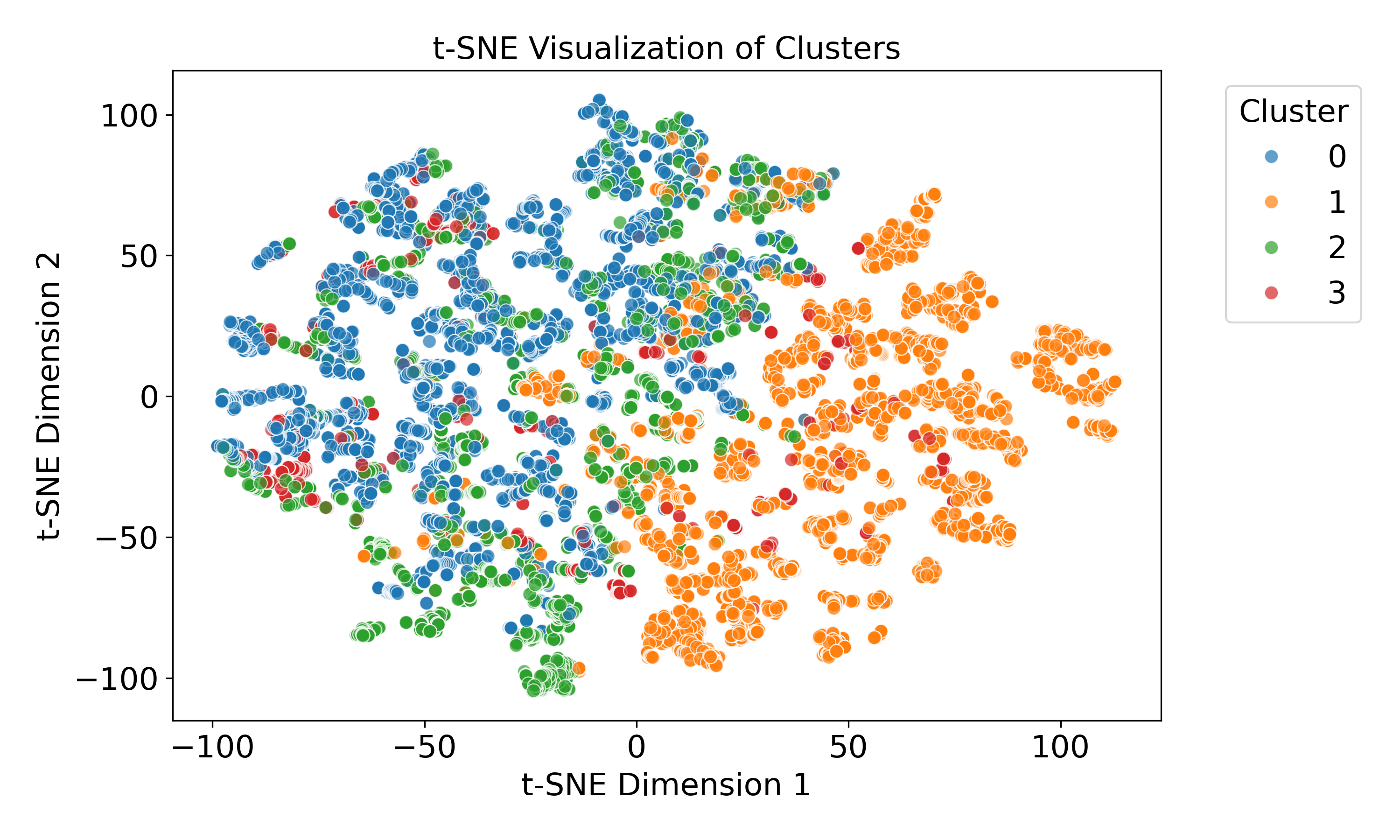}
        \caption*{(b) t-SNE projection with cluster labels}
    \end{minipage}%
    \caption{{Figure 5: (a) Cluster-wise elemental composition percentages highlighting the dominant chemical elements within each K-means cluster. The distributions reveal clear chemical themes, such as Fe/Co-rich metallic systems, oxide-based compositions enriched in O, Mn, and La, and Mn/Ni-containing intermetallics. (b) t-SNE projection of the high-dimensional descriptor space into two dimensions, with each point representing a material colored by its assigned cluster. The distinct separation—especially the well-isolated oxide-rich Cluster 1—demonstrates that the clustering captures meaningful compositional differences relevant to Curie temperature variation.} }
    \label{fig:tsne_cluster_composition_combined}
\end{figure*}

We then applied the Extra Trees algorithm independently to each cluster and observed that Clusters 1 and 3 exhibited significantly lower predictive performance, with R$^2$ values of approximately 0.58 and 0.38, respectively. Cluster 1 was found to be dominated by oxygen-rich compounds, particularly complex oxides, as illustrated in Figure~\ref{fig:tsne_cluster_composition_combined}. Examples of compounds in this cluster include, LaMnO$_3$, SrFeO$_3$, and CaMnO$_3$. From a physical standpoint, such oxide systems differ substantially from the metallic systems that dominate other clusters. Their magnetism is governed by mechanisms like superexchange, strong electron correlation, and structural distortions (e.g., Jahn–Teller effects), which are not well captured by simple compositional descriptors. This explains the degradation in model performance within this cluster.

Cluster 3 yields the lowest $R^2$ compared to other clusters, with the highest MAE and RMSE values among all clusters. This underperformance likely arises from both its small size and its unique chemical composition. Unlike other clusters dominated by transition metal oxides or alloys, Cluster 3 contains a high proportion of carbon, and phosphorus-rich compounds, including intermetallics and carbides. These materials often exhibit complex bonding environments and non-collinear magnetic structures, which are challenging to model using composition-only descriptors. The limited number of training samples in this chemically diverse space further hinders generalization.

Therefore, the underperformance of the model in Clusters 1 and 3 provides a strong motivation for incorporating structure-aware or quantum-level descriptors in future studies. Their distinct magnetic behaviors and modeling challenges highlight the limitations of current descriptor sets and the need for more physically grounded representations.

Consequently, excluding these clusters from our analysis improves overall model performance and this improves the value of $R^2$ to approximately 0.85. This decision is supported both statistically and physically and highlights the importance of domain knowledge when applying machine learning to datasets for magnetic materials. After removing Clusters 1 and 3 from the data, we obtain the following results from 5-fold cross validation:

\begin{table}[h]
\centering
\caption{Cross-validation (5-fold) results for the best-performing model.}
\begin{tabular}{|l|c|c|}
\hline
\textbf{Metric} & \textbf{Value} & \textbf{Mean $\pm$ Std} \\
\hline
MAE (K)   & 54 & 54 $\pm$ 2.5 \\
RMSE (K)  & 105 & 105 $\pm$ 4.0 \\
R\textsuperscript{2} & 0.85 & 0.85 $\pm$ 0.012 \\
\hline
\end{tabular}
\label{tab:crossval_results}
\end{table}

\subsection*{t-SNE Visualization of Clusters}
To further understand the separation and characteristics of clusters formed in the dataset, we employed t-SNE (t-distributed Stochastic Neighbor Embedding) to reduce the high-dimensional feature space to two dimensions. The resulting 2D visualization, shown in Figure~\ref{fig:tsne_cluster_composition_combined}, illustrates clear separation between chemically distinct groups.

Cluster 0 contains high concentrations of metallic elements such as iron (Fe), cobalt (Co), silicon (Si), and boron (B), reflecting typical magnetic intermetallics.
Cluster 1 is dominated by complex oxides, rich in oxygen (O), manganese (Mn), lanthanum (La), and strontium (Sr)—a chemical signature common in perovskite and layered oxide systems. 
Cluster 2 is enriched in manganese (Mn), iron (Fe), gallium (Ga), and nickel (Ni), resembling Heusler-type alloys.
Cluster 3 includes mainly carbon (C), iron (Fe), and Cobalt (Co), and represents simpler binary and ternary compounds. 

These clusters not only form distinct regions in t-SNE space but also exhibit coherent chemical themes, validating the clustering strategy and emphasizing the influence of composition on Curie temperature behavior.

\begin{figure*}[t]
    \centering
    \begin{minipage}[t]{0.48\linewidth}
        \centering
        \includegraphics[width=\linewidth]{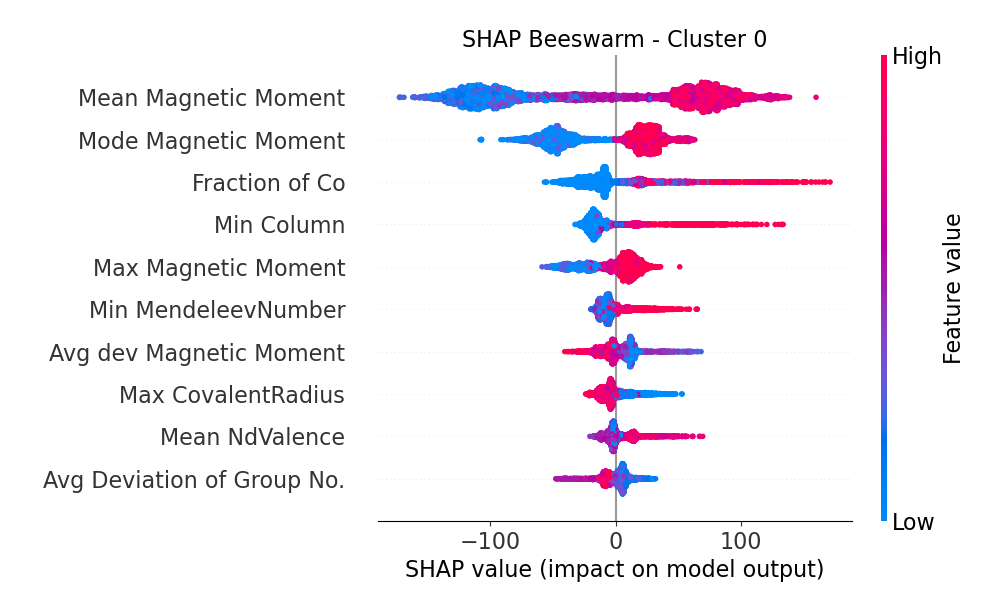}
        \caption*{Cluster 0}
    \end{minipage}\hfill
    \begin{minipage}[t]{0.48\linewidth}
        \centering
        \includegraphics[width=\linewidth]{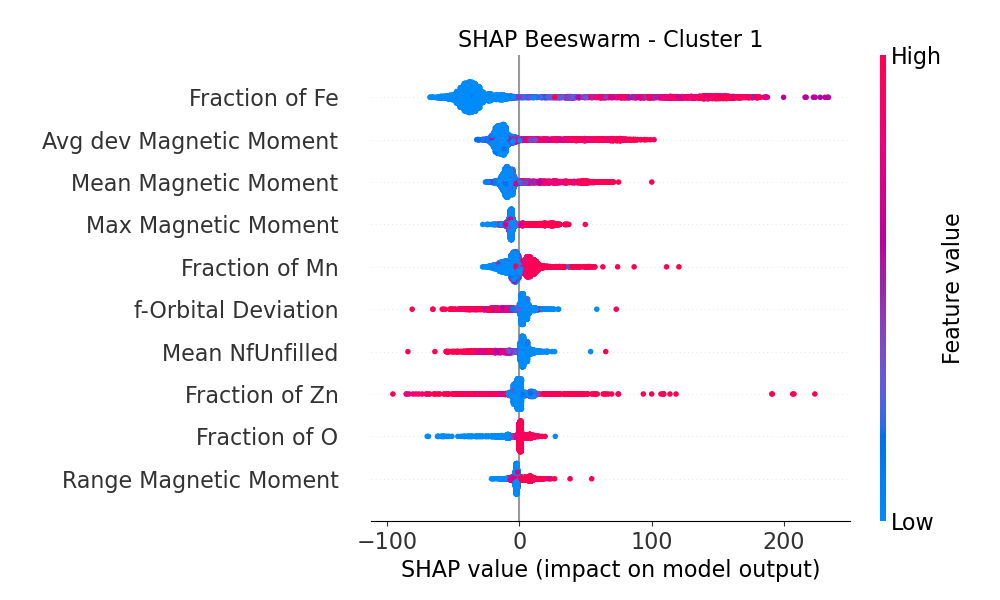}
        \caption*{Cluster 1}
    \end{minipage}

    \vspace{0.5cm}

    \begin{minipage}[t]{0.48\linewidth}
        \centering
        \includegraphics[width=\linewidth]{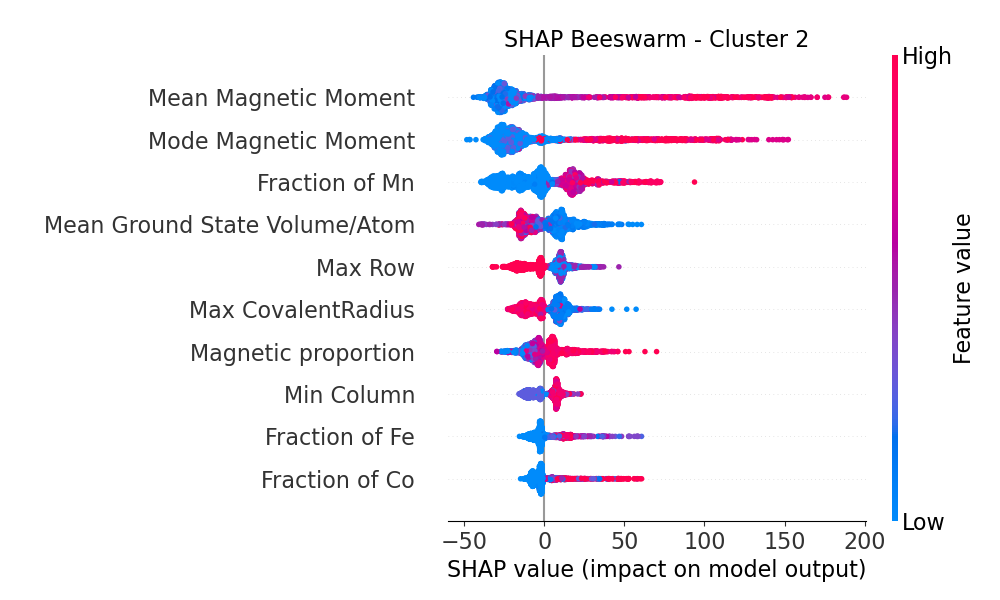}
        \caption*{Cluster 2}
    \end{minipage}\hfill
    \begin{minipage}[t]{0.48\linewidth}
        \centering
        \includegraphics[width=\linewidth]{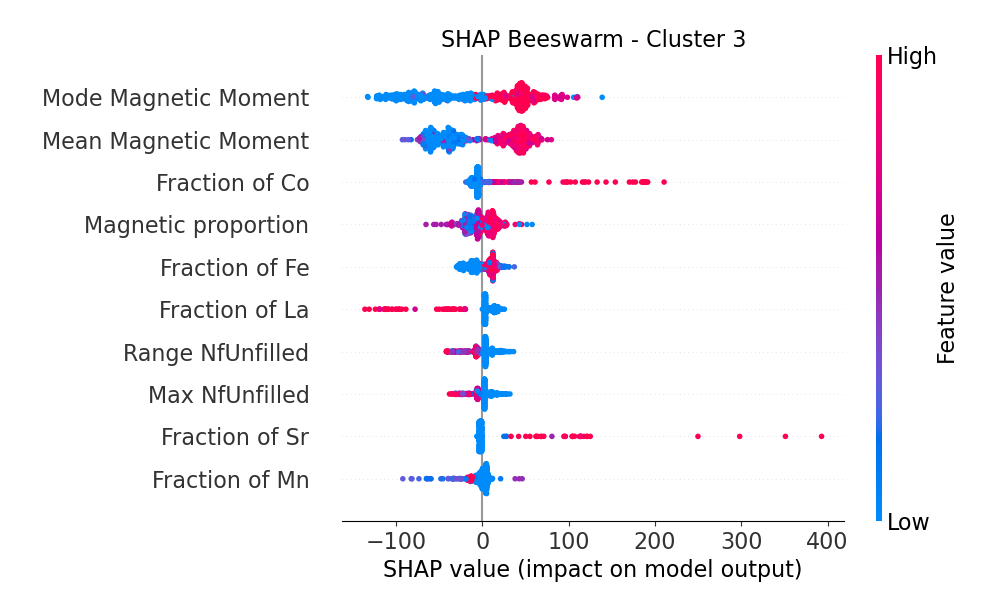}
        \caption*{Cluster 3}
    \end{minipage}
    
    \caption{{SHAP beeswarm plots for the four K-means clusters, showing how feature values contribute to Curie temperature predictions within each chemical subgroup. Cluster 0 (top left) contains Fe/Co-rich materials where magnetic-moment descriptors are particularly strong contributors. Cluster 1 (top right) includes oxide-based compounds, where magnetic-moment features remain important but orbital-deviation and magnetic-variance descriptors gain greater influence. Cluster 2 (bottom left), composed of Mn/Ni intermetallics, is primarily driven by magnetic-moment and atomic-size descriptors. Cluster 3 (bottom right) is a smaller, compositionally diverse group with a relatively broader SHAP contribution.}}
    \label{fig:shap_all_clusters}
\end{figure*}

\section{Cluster-wise SHAP Analysis}

To further enhance the interpretability of our machine learning models for Curie temperature ($T_c$), we applied the SHAP analysis for each of the four clusters identified via KMeans. By isolating materials into chemically similar groups, we can uncover subtle relationships and also better understand which descriptors govern $T_c$ in different regimes.

{Cluster 0}, which is the largest cluster (4,744 materials), contains a significant number of Fe-, Co-, and rare-earth-based compounds. The model achieves strong performance with MAE = 86.83~K and $R^2 = 0.80$. As shown in top left panel of Figure~\ref{fig:shap_all_clusters}, the most influential features include the {mean magnetic moment}, {mode magnetic moment}, and {fraction of Co}, all positively correlated with $T_c$. These results reflect known physical principles — high magnetic moments enhance exchange interactions and stabilize ferromagnetic ordering. Additional features such as {minimum Mendeleev number} and {mode column} suggest that periodic trends play a secondary yet non-trivial role.

{Cluster 1} (4,826 materials) includes many oxide-based perovskites rich in Mn, La, and O. The model performance is lower (MAE = 100.21~K, $R^2 = 0.58$), likely due to the structural and chemical complexity of these materials. Figure~\ref{fig:shap_all_clusters} (top right plot) reveals that {fraction of Fe}, {average deviation of magnetic moment}, and {f-orbital deviation} are key predictive features. The prominence of orbital and magnetic variance descriptors suggests the presence of competing magnetic mechanisms such as double exchange and Jahn-Teller distortions. These subtleties may explain the slightly reduced predictive power compared to Cluster 0.

Cluster 2 (2,730 materials) consists of Mn- and Ni-based intermetallics, including Heusler-like alloys. Despite compositional richness, the model achieves good performance (MAE = 88.68~K, $R^2 = 0.73$). In Figure~\ref{fig:shap_all_clusters} (bottom left plot), features such as {mean magnetic moment}, {fraction of Mn}, and {maximum covalent radius} dominate. This aligns with the tunability of intermetallic compounds, where structural and electronic changes can significantly influence magnetic ordering. The importance of spatial descriptors further highlights the role of atomic packing and size effects in modulating $T_c$.

Cluster 3 is the smallest cluster (760 materials) and exhibits the poorest performance (MAE = 138.88~K, $R^2 = 0.38$). This may be due to its small size and its unique chemical composition. As seen in Figure~\ref{fig:shap_all_clusters}  (bottom right plot), while magnetic descriptors still lead, there is greater importance given to features such as {La fraction}, {Sr fraction}, and {f-orbital deviation}. These may hint at more subtle magnetic phenomena that our current descriptor set cannot fully capture. The high variance in SHAP values also suggests that no single feature consistently governs $T_c$ across the cluster, pointing to the need for more tailored descriptors or domain-specific models.

SHAP analysis across clusters confirms that magnetic moment descriptors are universally important predictors of Curie temperature. However, the prominence of different supporting features (e.g., orbital variances, group numbers, atomic radii) shifts based on the chemical and structural nature of the materials within each cluster. This highlights that even with a shared  target, distinct materials demand distinct explanatory regimes.

We conclude that cluster-specific modeling not only improves performance but also enriches physical insight. Future work may benefit from hybrid clustering methods (e.g., hierarchical or domain-guided) and the integration of additional physics-based descriptors tailored to each family. It is also worth noting that while SHAP helps explain model behavior, it doesn't always align with physical intuition, especially for complex oxides. This is expected, as SHAP reflects correlations based on available descriptors, which may miss deeper physics. Its insights are valuable, but approximate.

\section{Conclusion}
This study demonstrates that explainable machine learning can serve as a powerful tool for materials property prediction, particularly for estimating Curie temperatures using composition-based descriptors. We employed various machine learning models and found that, compared to other models, the optimized Extra Trees Regressor demonstrated robust predictive performance for Curie temperature estimation. 

Furthermore, by leveraging explainable machine learning, we uncover feature-driven insights into the compositional factors that influence ferromagnetism in inorganic materials. The use of SHAP analysis allowed us to interpret model outputs in a physically meaningful way.   Additionally, the application of clustering and t-SNE visualization revealed distinct patterns and subgroups within the dataset.

Our framework demonstrates that incorporating explainable AI not only improves transparency but also further strengthens the scientific value of machine learning models in materials science. This approach can be readily extended to other property prediction tasks across the domain.

\section{Acknowledgments}
We would like to thank Babar Shabbir for useful discussions.

The code and data for this project are available on GitHub: \url{https://github.com/dradeelajaib/Curie-Prediction}

\bibliographystyle{plainnat}

\end{document}